\documentclass{kluwer}    

\newdisplay{guess}{Conjecture}

\usepackage{psfig}

\begin{document}                                                                                   
\begin{article}
\begin{opening}         
\title{Prebiotic Homochirality as a Critical Phenomenon} 
\author{Marcelo \surname{Gleiser} and Joel \surname{Thorarinson}}  
\runningauthor{Gleiser \& Thorarinson}
\runningtitle{Prebiotic Homochirality as a Critical Phenomenon}
\institute{Department of Physics and Astronomy, Dartmouth College,
Hanover, NH 03755, USA}
\date{November 20, 2005}

\begin{abstract}
The development of prebiotic homochirality on early-Earth or another planetary
platform may be viewed as a critical phenomenon. It is shown, in the context 
of spatio-temporal polymerization reaction networks, that environmental effects
--- be them temperature surges or other external disruptions --- may destroy
any net chirality previously produced. In order to understand the emergence of
prebiotic homochirality it is important to model the coupling of polymerization
reaction networks to different planetary environments.
\end{abstract}

\end{opening}           

\section{Introduction}  
                    
According to the prebiotic soup hypothesis \cite{Bada00},  the early Earth had
the supply of organic compounds needed to jump-start polymerization  reactions
that, through gradual complexification, led to the first
biochemical networks displaying some of the characteristics attributed to
life,  such as metabolic activity and replication. Although the road is
still obscure \cite{Orgel98}, the situation is not
all bleak. In 1953, Stanley Miller
simulated tentative conditions of early-Earth in the laboratory to  obtain
amino acids from simple chemical compounds.
That same year, Frank
\cite{Frank53} proposed that auto-catalytic polymerization could explain the
emergence of biomolecular homochirality, a clear signature of terrestrial and,
possibly, all life \cite{Bonner}: terrestrial amino acids belonging to proteins
are overwhelmingly left-handed, while sugars are right-handed.

If a bottom-up approach to the early development of life is adopted,  the
homochirality of life's biochemistry must have emerged dynamically, as 
reactions among the simplest molecular building blocks occurred with high
enough yield. Alternatively, one may assume that, somehow, only monomers 
of a single chirality were present in the prebiotic soup: they were
made that way or brought here during the intense bombardment of
Earth's infancy, that lasted to about 3.8 Gyr ago \cite{Gomes05}. We would,
however, still need to understand how homochirality developed elsewhere in the
cosmos and not here, and whether it developed with the same chiral bias in more
than one place.

Here, we consider the homochirality of life as an
emergent process that took place on early-Earth and, possibly,
other planetary platforms. As a starting point, we use the
reaction-network model proposed by Sandars \cite{Sandars03},  which includes
enantiometric cross-inhibition. As shown in the
interesting work of Brandenburg and Multam\"aki \cite{BM}  (BM), Sandar's
polymerization reaction network can be reduced to an effective spatio-temporal
mean-field model, where the order parameter is the chiral asymmetry between
left and right-handed polymers. To this, we add the effects of an external
environment, showing that they can be crucial in the final determination of the
net value of enantiometric excess, if any.

\section{Modeling Polymerization}

Sandars proposed the following polymerization reactions \cite{Sandars03}:
$L_n + L_1 \stackrel{2k_S}{\longrightarrow} L_{n+1}$;
$L_n + R_1 \stackrel{2k_I}{\longrightarrow} L_nR_1$;
$L_1 + L_nR_1 \stackrel{k_S}{\longrightarrow} L_{n+1}R_1$; and,
$R_1 + L_nR_1 \stackrel{k_I}{\longrightarrow} R_1L_nR_1$,
where $k_S$ ($k_I$) are the reaction rates for adding monomers of the same
(opposite) chirality to a given chain. The network is completed by adding the
four opposite reactions, that is, by interchanging $L\leftrightarrow R$, and by
adding a substrate $S$ from which both left and right-handed monomers emerge,
$S\stackrel{k_CC_L}{\longrightarrow} L_1$ and
$S\stackrel{k_CC_R}{\longrightarrow} R_1$,
where $C_{L(R)}$ determine the enzymatic enhancement of left and right-handed
monomers, which is not known. We follow Sandars \shortcite{Sandars03} and choose
$C_L = [L_N]$, $C_R=[R_N]$. As remarked in BM, it is
possible to truncate the system to $N=2$ and still maintain the essential
aspects of the dynamics leading to homochiralization. This 
allows us to model the reaction network as a mean-field
theory exhibiting spontaneous chiral symmetry breaking \cite{Weinberg96}. Our
approach blends the work of BM
with the pioneering work of Kondepudi and Nelson, where the
reaction network was coupled to time-dependent external effects \cite{KN85}:
chirality evolves spatio-temporally in contact with an
environment.

The equations can be simplified by assuming that the rate of
change of $[L_2]$ and $[R_2]$ is much slower than that of $[L_1]$ and $[R_1]$.
The same for the substrate $[S]$, so that $d[S]/dt = Q-(Q_L+Q_R)
\simeq 0$, where $Q_L$ and $Q_R$ are the source terms for monomers generated
from the substrate $[S]$:
$Q_L=k_C[S](pC_L + qC_R)$, and $Q_R=k_C[S](pC_R + qC_L)~$ \cite{Haken83}. The
constants $p=\frac{1}{2}(1+f)$ and $q=\frac{1}{2}(1-f)$ are given in terms of
the fidelity of enzymatic reactions $f$, an adjustable parameter. As
demonstrated by Kondepudi and Nelson \cite{KN83} and many others
\cite{Sandars03, WC05, SH04, GR05, AG93}, auto-catalytic
networks show a tendency to bifurcate toward homochirality for values
above critical, $f_c$. The specific value of $f_c$ is model-dependent.

Under the above assumptions, and introducing the dimensionless
symmetric and asymmetric variables,
${\cal S}\equiv X+Y$ and ${\cal A}\equiv X-Y$, where
$X\equiv [L_1](2k_S/Q)^{1/2}$ and $Y\equiv [R_1](2k_S/Q)^{1/2}$, 
BM have shown that the polymerization equations reduce to
\begin{eqnarray}
\lambda_0^{-1}\frac{d{\cal S}}{dt}&=&1-S^2\\
\lambda_0^{-1}\frac{d{\cal A}}{dt}&=&2f\frac{{\cal S}{\cal A}}{{\cal S}^2+
{\cal A}^2} - {\cal S}{\cal A}~,
\end{eqnarray}
where $\lambda_0\equiv (2k_SQ)^{1/2}$, with
dimension of inverse time. 
${\cal S}=1$ is a fixed point: the system will tend towards this value at
time-scales of order $\lambda_0$. With ${\cal S}=1$, the equation for the
chiral asymmetry has fixed points at ${\cal A}=0,~\pm\sqrt{2f-1}$, as pointed
out in BM. An enantiometric excess is only possible if $f>f_c=1/2$.

\section{Introducing Environmental Effects}

We model the external environment via a stochastic spatio-temporal
Langevin equation, rewriting the equations above as
\begin{eqnarray}
\label{aeq}
\lambda_0^{-1}\left(\frac{d{\cal S}}{dt} -k\nabla^2{\cal S}\right)&=&
1 - S^2 + \lambda_0^{-1}\xi({\bf x},t)\\
\lambda_0^{-1}\left( \frac{d{\cal A}}{dt}-k\nabla^2{\cal A}\right)&=&
2f\frac{{\cal S}{\cal A}}{{\cal S}^2+
{\cal A}^2} - {\cal S}{\cal A} + \lambda_0^{-1}\xi({\bf x},t)~,
\end{eqnarray}
where $k$ is the diffusion constant and $\xi$ represent white noise with
zero mean and a two-point correlation function given by
$\langle \xi({\bf x'},t')\xi({\bf x},t)\rangle = a^2\delta({\bf x'}-{\bf x})
\delta(t'-t)$,
and $a^2$ measures the strength of the external influence. For example, in
mean-field models of phase transitions, it is common to write $a^2=2\gamma
k_BT$, where $k_B$ is Boltzmann's constant, $T$ is the temperature, and
$\gamma$ is a viscosity coefficient. The equations 
can be made dimensionless by introducing 
$t_0=\lambda_0t$ and $x_0=\sqrt{\lambda_0/k}x$, which determine the typical 
spatio-temporal scales in the system. The noise amplitude scales as 
$a^2_0\rightarrow \lambda_0^{-1}(\lambda_0/k)^{d/2}a^2$, where $d$ is the
number of spatial dimensions. Using as nominal values for the
parameters, $k_S\sim 10^{-25}$cm$^3$s$^{-1}$, $Q\sim 10^{15}$cm$^{-3}$s$^{-1}$, the
diffusivity of water $k=10^{-9}$m$^2$s$^{-1}$, we obtain, 
$t\simeq (7\times 10^4$s)$t_0$ and $x\simeq (1$cm)$x_0$.

As in BM [see also \cite{SH04, GR05}], the
concentrations are spatially-dependent quantities. This implicitly assumes
that it is possible to define an effective correlation volume within which the
value of the chiral asymmetry ${\cal A}$ is fairly
homogeneous. (We take ${\cal S}=1$.) Using well-known
results from the mean-field theory of phase transitions \cite{Landau80}, we can
easily compute the correlation length. From the equation
for ${\cal A}$, we obtain an effective potential $V({\cal A})$, 
\begin{equation}
\label{V}
V({\cal A}) = \frac{{\cal A}^2}{2} - f\ln\left [{\cal A}^2+1\right ]~.
\end{equation}
For $f<1/2$, $V({\cal A})$ has a typical double-well shape, 
with minima at the fixed points 
${\cal A}_{\pm}=\sqrt{2f-1}$. The correlation length,
$\xi$, is given by
$\xi^{-2}({\cal A}_{\rm min}) = V^{\prime\prime}({\cal A}_{\rm min})$,
where ${\cal A}_{\rm min}$ denotes a minimum of the potential. For the fixed
points, we get, $\xi^2({\cal A}_{\pm}) = f/(2f-1)$.  At
$f_c=1/2$ the correlation length diverges, as it should for a critical point. 
However, the noise
parameter $a$ also controls the behavior of the system. Indeed, even if $f=1$,
an enantiometric excess may not develop if $a$ is above a critical value $a_c$.
In analogy with ferromagnets, where above a critical temperature the net
magnetization is zero, one may say that above $a_c$ the stochastic forcing due
to the external environment overwhelms any local excess of $L$ over $R$
enantiometers within  a domain of correlation volume $V_{\xi}\sim \xi^d$:
racemization is achieved at large scales and chiral symmetry is restored
throughout space. 

\section{Numerical Results: Critical Point for Homochirality}

Salam \cite{Salam91} suggested that there should be a critical temperature
$T_c$ above which any net homochirality is destroyed. However, he conceded that
calculating $T_c$ would be quite challenging using the  electroweak theory of
particle physics. Here, we chose a different route which, we believe, will
allow us to explore the qualitative aspects of the problem more effectively:
the noise amplitude $a$ may represent a sudden increase in temperature and/or
pressure due  to a meteoritic impact or volcanic eruption, or, possibly, due to
a source of circularly-polarized ultraviolet light \cite{Lucas05}.

\begin{figure}[t]
	\centerline{\psfig{figure=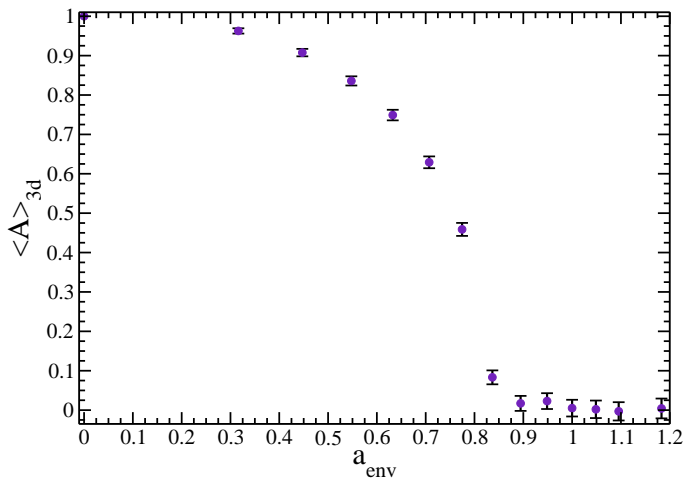,width=4.in,angle=270}}
	\caption{Average enantiometric excess versus ``temperature'' 
	in three dimensions.}
\label{tcritfig}
\end{figure}

The equation dictating the evolution of the enantiometric excess ${\cal A}$,
eq. \ref{aeq}, was solved with a finite-difference method in a $1024^2$ grid
and a $100^3$ grid with $\delta t=10^{-3}$ and $\delta x=0.2$,  and periodic
boundary conditions. In 2d, this corresponds to simulating a shallow pool with
linear dimensions of $\ell\sim 200$cm.  We prepared the system initially in a
chirally pure phase, which we chose to be  $\langle {\cal A}\rangle (t=0)=1$.
The equation is then solved for different values of the external noise, $a$. As
can be seen in Figure \ref{tcritfig}, for $a^2>a_c^2\simeq 0.65
(k/\lambda_0)^{3/2}$, $\langle {\cal A}\rangle \rightarrow 0$, that is, the
system becomes racemized. $\langle {\cal A}\rangle$ approaches a constant for
large times, indicating that the reaction network reaches equilibrium with the
environment. The results are ensemble averaged. For $d=2$, $a^2_c\simeq 1.15
(k/\lambda_0)$.

We can describe the environmental impact on
homochirality at the microscopic level by introducing the ``Ginzburg
criterion'' familiar of phase transitions \cite{Landau80}.  Consider a
correlation volume with $\langle {\cal A}\rangle =1$ (or $-1$). What is the
energy barrier ($E_G$) to flip half the molecules in the volume so that
$\langle {\cal A}\rangle \rightarrow 0$?  If $N_{\xi}$ is the number of
molecules in a correlation volume, $E_G = (N_{\xi}/2)E_f$, where $E_f$ is the
energy to flip one molecule. The Ginzburg criterion says that this energy is
$E_G\simeq V_{\xi}\Delta V$, where $\Delta V = |V(0)-V(\pm 1)|$.
Comparing the two expressions we obtain,
$E_f = 2\Delta V(V_{\xi}/N_{\rm \xi})$.
From Equation \ref{V}, $|V(0)-V(\pm 1)|=0.193$. Now, $V_{\xi} \simeq 4\xi^3=4
(k/\lambda_0)^{3/2}$. (We set $f=1$.) Using for the microscopic spatial scale 
$\xi_{\rm micro} \simeq (Q/k_S)^{-1/6}$, and that $N_{\xi}\simeq (\xi/\xi_{\rm
micro})^3$, we obtain $E_f \simeq 1.5\times 10^{-26}$m$^3$. 
[The energy has dimensions of (length)$^d$.] To
complete the argument, we use that the  critical ``environmental'' energy to
restore the chiral symmetry was obtained numerically to be (cf. Figure
\ref{tcritfig}),  $E_{\rm env}\simeq 0.65(k/\lambda_0)^{3/2}=0.65\times
10^{-6}$m$^3$. We thus obtain the ratio, $E_f/E_{\rm env}|\simeq 2.3\times
10^{-20}$.

It is tempting to compare this result with possible sources of homochirality.
For example, weak neutral currents are expected to induce an excess at room
temperature of $g=\Delta E/k_BT \sim 10^{-17}$ \cite{KN83}.  In the language of
the present work, they would induce a tilt in the potential $V({\cal A})$
proportional to $g$. Thus, within the violent environment of prebiotic Earth,
effects from such sources,  even if cumulative, would be negligible: any
accumulated excess could be easily wiped out by an external disturbance.
Further work along these lines is in progress.


\end{article}
\end{document}